\documentclass[12pt]{article}
\usepackage{graphicx}
\usepackage{amssymb,amsmath,amsfonts,palatino,amsthm}
\usepackage{amssymb}
\usepackage{epstopdf}
\usepackage{color}
\newcommand{\beq}{\begin{equation}}
\newcommand{\eeq}{\end{equation}}

\newcommand{\f}{\begin{equation}}
\newcommand{\ff}{\end{equation}}

\begin{document}

\title{Spin foam models as energetic causal sets \\}
\author{Marina Cort\^{e}s${}^{1,2,3}$ and Lee Smolin${}^{1}$
\\
\\
Perimeter Institute for Theoretical Physics${}^{1}$\\
31 Caroline Street North, Waterloo, Ontario N2J 2Y5, Canada
\\
\\
Institute for Astronomy, University of Edinburgh ${}^{2}$\\
Blackford Hill, Edinburgh EH9 3HJ, United Kingdom
\\
\\
Centro de Astronomia e Astrof'sica da Universidade de Lisboa${}^{3}$\\
Faculdade de Ci\^encias, Ed. C8, Campo Grande, 1769-016 Lisboa, Portugal}
\date{\today}
\maketitle

\begin{abstract}

Energetic causal sets are causal sets endowed by a flow of energy-momentum between causally related events.  These incorporate a novel mechanism for the emergence of space-time from causal relations \cite{ECS1,ECS2}.   Here we construct a spin foam model which is also an energetic causal set model.  This model is closely related to the model introduced in parallel by Wolfgang Wieland in Ref.~\cite{WWnew}.  What makes a spin foam model also an energetic causal set is Wieland's identification of new momenta, conserved at events (or four-simplices), whose norms are not mass, but the volume of tetrahedra. This realizes the torsion constraints, which are missing in previous spin foam models, and are needed to relate the connection dynamics to those of the metric, as in general relativity.  
This identification makes it possible to apply the new mechanism for the emergence of space-time to a spin foam model.  

Our formulation also makes use of Markopoulou's causal formulation of spin foams\cite{fotini-dual}.  These  are generated by evolving spin networks with dual Pachner moves. This endows the spin foam history with causal structure given by a partial ordering of the events which are dual to four-simplices.
\end{abstract}

\newpage

\tableofcontents

\section{Introduction}

The notions of causality and causal structure is central to special and general relativity, yet it is controversial whether it plays a fundamental role in quantum gravity. They play a prominent role in some approaches to quantum gravity, including causal sets, causal dynamical triangulations, relative locality and twistor theory-to mention a disparate group. Most studies of spin foam models, on the other hand, de-emphasize the role of causal structure.  However there are exceptions.
In \cite{fotini-dual} spin foams generated by evolving a dual spin network with Pachner moves, were given a causal structure by establishing a partial order between events.  In \cite{FS_SF} a more global construction was proposed, similar to that explored in causal dynamical triangulation models.   Both these formulations, by providing a framework in which dual spin networks were evolved causally, married loop quantum gravity to causal set models\footnote{Other ways to impose causal structures on spin foam models were studied in\cite{other}.  Another related approach is described in
\cite{other2}.}.

In \cite{ECS1,ECS2} we introduced a version of causal sets in which  events are endowed with momentum and energy transmitted along causal links, and playing a role in establishing the causal structure.  We call these energetic causal sets.  One feature they have which distinguish them from bare causal sets is that they incorporate a novel mechanism for the emergence of a classical space-time in the semiclassical limit\cite{ECS1,ECS2}.  In this mechanism, space-time co-ordinates of events, which are absent when the model is first formulated, arise as lagrange multipliers used in the expression of the constraints that enforce the conservation of energy-momenta at events.  

In this paper we establish that a new kind of spin foam model, introduced by Wieland \cite{WWnew},  can be understood as an energetic causal set model, with additional structure.  We achieve this correspondence in three steps.  The first is to build a spin foam as a causally evolved spin-network following \cite{fotini-dual}.  This construction is reviewed in the next section.  The second step is to add holonomy and flux variables appropriate to the dynamics of loop quantum gravity.  The third step is to follow  Wieland \cite{WWnew} in adding to the spin foam model a conservation law for a new kind of four-momentum, assigned to tetrahedra.  The four momenta of a tetrahedron is normal to it, while its norm is proportional to the tetrahedra's volume.  These momenta are conserved at events, which are dual to four simplices, and the conservation enforces the closure of the four simplex, as made of its five constituent tetrahedra. 

As shown by Wieland\cite{WWnew}, this conservation law realizes the imposition of the constraint that the torsion of the space-time connection vanishes.  This is necessary if the dynamical connection is to carry information about the metric and frame fields, necessary to turn the first order dynamic of constrained $BF$ theories into the metric dynamics of general relativity.  

This conservation law also makes possible the identification of the causal set model with an energetic causal set.  This is the main claim of this paper.  This we expect will be useful as it gives a new route to the emergence of classical space-time and general relativity from the semiclassical limit of a spin foam model.

The spin foam model we describe here is then very closely related to that presented by Wieland in \cite{WWnew}.  The main difference is that we work with an action that is purely discrete, whereas Wieland uses an action continuous in a time parameter that parameterizes the edges of faces of the simplicial complex.  This formulation realizes a beautiful Hamiltonian structure.  For some cases, the continuous of the action, and the related symplectic structures, may be derived as a limit of our discrete action, along the lines of the derivation of the free particle action from the discrete action in energetic causal sets in \cite{ECS1,ECS2}.   


In this paper we aim to be pedagogical and so we assume no prior knowledge of causal sets or spin foam models.  Indeed, the technical complexities of spin foam models can be postponed till the last stage of the construction.  
In the next section we recall the construction of causally evolving dual spin networks from \cite{fotini-dual}.  In section 3 we add momenta and turn these into energetic causal sets.  In section 3 we add additional degrees of freedom that code geometrical information and so turn the model into a spin network model, closely related to that of \cite{WWnew}.  It is at this last step that the identification of mass with volume is made.

\section{Recalling dual spin network causal evolution}

The marriage of loop quantum gravity and causal sets, leading to a formulation of causal spin foams, was proposed by Markopoulou\cite{fotini-dual}.  We'll describe this first for 2+1d, where it is simpler to analyze, then extend to 3+1d.

\subsection{2+1d Spin Networks}

\begin{figure}[t!]
\centering
\includegraphics[width= 0.7\textwidth]{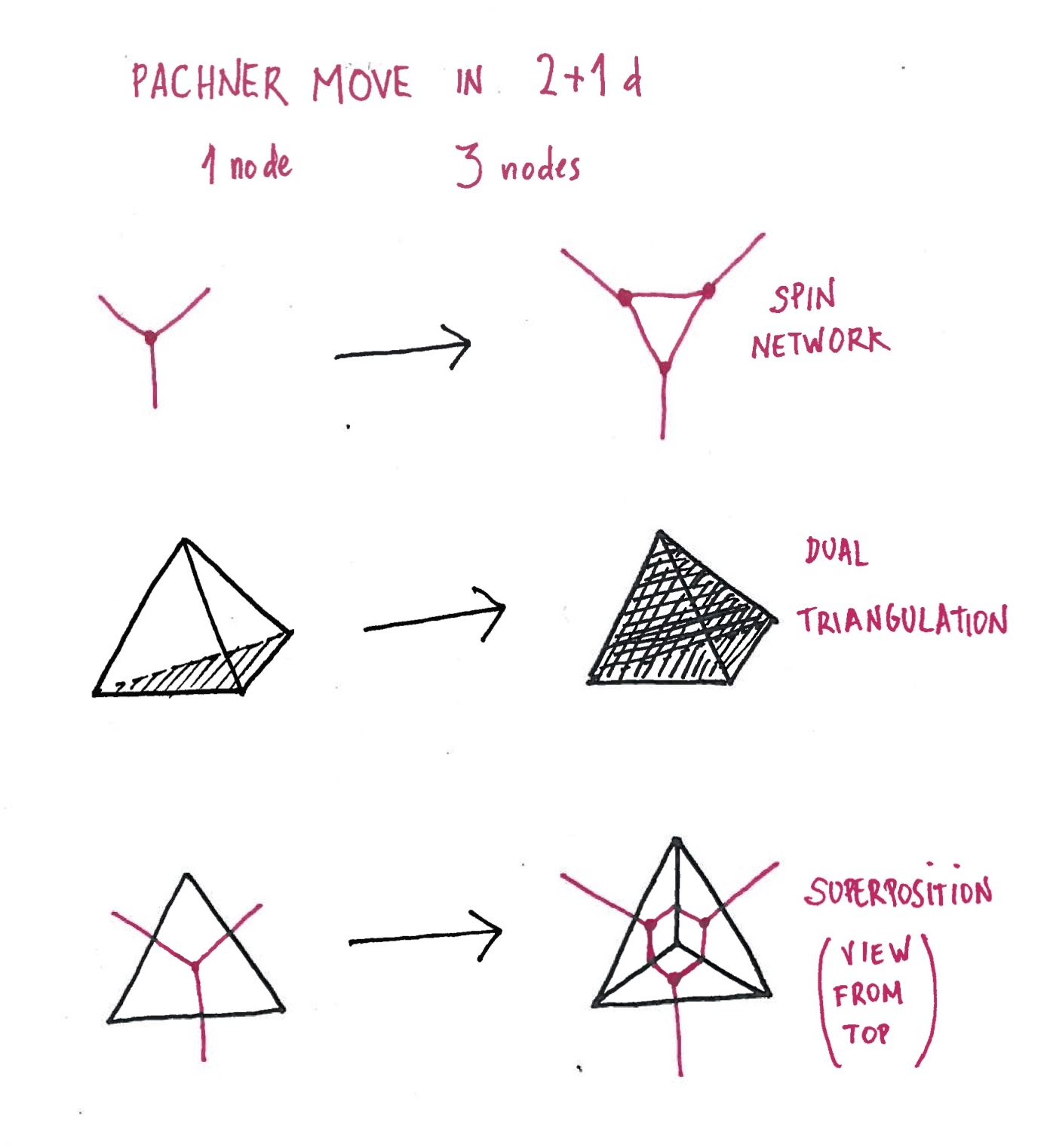}
\caption{$1 \rightarrow 3$ Pachner moves in 2+1d, in the dual spin network/dynamical triangulation representation. 
{\bf(Upper panel)} Spin network move: one node becomes three nodes. {\bf(Middle Panel)} $1 \rightarrow 3$ move in triangulations: the lower triangle becomes the upper three triangles. {\bf(Bottom panel)} View from top: both triangulation and dual spin network superposed}
\label{pachner1_3} 
\end{figure}

\begin{figure}[t!]
\centering
\includegraphics[width= 0.7\textwidth]{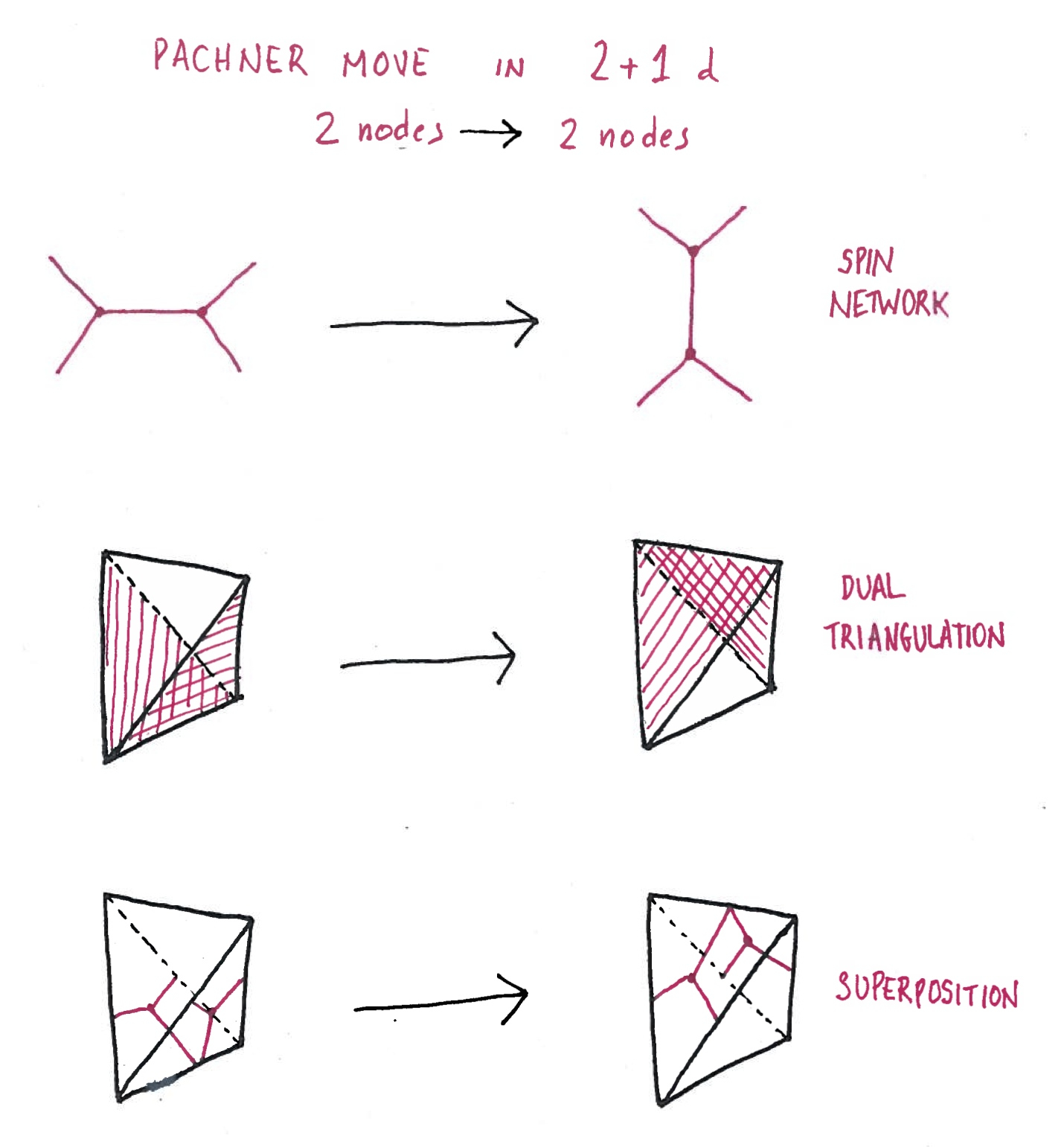}
\caption{$2 \rightarrow 2$ Pachner moves in 2+1d, in the dual spin network/dynamical triangulation representation.}
\label{pachner2_2} 
\end{figure}

\begin{itemize}

\item Causally evolving spin networks are constructed from evolving states by one of a set of local evolution moves. In 2+1d a state is represented by a triangulation of a space-like surface. 
An evolution move is a discrete time step called Pachner move.  Each Pachner move performed on the spatial slicing corresponds to an event.

\item Each triangle in the spatial triangulation represents a locally flat piece of 2d space.  The triangulation is dual to a three-valent spin network $\Gamma_{i}$ embedded in a topological two manifold $\Sigma$. The center of each triangle is dual to a node in the spin network,  and labeled by intertwiners. The sides of each triangle are dual to edges in the spin network and labeled by $SU(2)$ spins.

\item From this triangulation we evolve to the next state by adding tetrahedra on top of it.  There are different kids of moves, each represented by a way to cover one, two or three adjacent triangles with the faces of the tetrahedra. For example, a so called $1 \rightarrow 3$ move is made  by adding one more point to the future of a given triangle, which creates a tetrahedron.  The initial triangle makes up the bottom (i.e. past) side of the tetrahedra.  This triangle is now replaced by the three new triangles making up the top, or future, side of the tetrahedron.  This tetrahedron represents the Pachner move and so generates the time step.  

The tetrahedron is formed by 4 glued triangles, part of these in the current spatial slice, the past, and part of these in the new spatial slice, the future. Splitting the 4 triangles in the tetrahedron between the past and future slices gives origin to different Pachner moves, and in 2+1d there are different 3 possibilities

\item In 2+1d the available Pachner moves are $1 \rightarrow 3$ triangles, $2 \rightarrow 2$, and $3 \rightarrow 1$.
If the tetrahedron is placed on top of one triangle in the current triangulation then that triangle is in the past slice and the three remaining triangles become part of the future triangulation, forming a $1 \rightarrow 3$ move, which we show in Figure~\ref{pachner1_3} in the dual spin foam/ dynamical triangulation representation. 
If it's placed on top of two adjacent triangles in the current triangulation, then the two complimentary triangles in the tetrahedron become part of the new representation, forming a $2 \rightarrow 2$ move, shown in Figure~\ref{pachner2_2}. Finally, if it's placed on top of three adjacent triangles in the existing triangulation, the remaining triangle becomes part of the new triangulation forming a $3 \rightarrow 1$ move. This is just the reverse of the $1 \rightarrow 3$ of Figure~\ref{pachner1_3}.

\item The Pachner moves are repeated many times over, creating a causal spin foam $SF$. In the language of ECS introduced in Section~\ref{ecs_intro} the Pachner moves represent events, $V_I$. Each tetrahedron $V_I$ is an event.

\item The resulting three dimensional simplicial complex is made from the events, which have the structure of a causal set.  Two events $V_I$ and $V_J$ have an immediate causal link, $L_{IJ}$ if a triangle in the future set of $I$ is also in the past set of $J$. Causal links between events are denoted edges. Edges represent time-like evolution and have a unique orientation  towards the future\footnote{Note that this is different from the model of \cite{WWnew} where there is no causal ordering.}. The causal link  $L_{IJ}$ can then contain several triangles, making a chain. We say event $K$ is to the future of event $I$, $J>I$  if there is a chain of  future pointing causal links beginning on $I$ and ending on $K$.  

\item A Pachner move represents the transition amplitude for an event to take place. This includes the generation of spins and intertwiners for the new edges and nodes introduced in the evolution move, as well as a choice of which of the available Pachner moves takes place.  This issue of identifying and attributing transition amplitudes to all the available Pachner moves will be addressed in future work, and is outside of the scope of the current work, which is purely quantum mechanical.

\item{} The edges and faces of the triangles are all space like.  Dual to each triangle, $T$  is a time-like link, $l_\tau$ connecting two events which contain $T$ as part of the future or past set.  

\item{} The causal network may include multiple time-like links between two causally related events. 

\item{} Except for the initial triangulation, every triangle is uniquely in the future set of one tetrahedron.   
Except for the final triangulation, every triangle is uniquely in the past set of one tetrahedron.  

\end{itemize}

%

\subsection{3+1d spin networks}

\begin{figure}[t!]
\centering
\includegraphics[width= 0.5\textwidth]{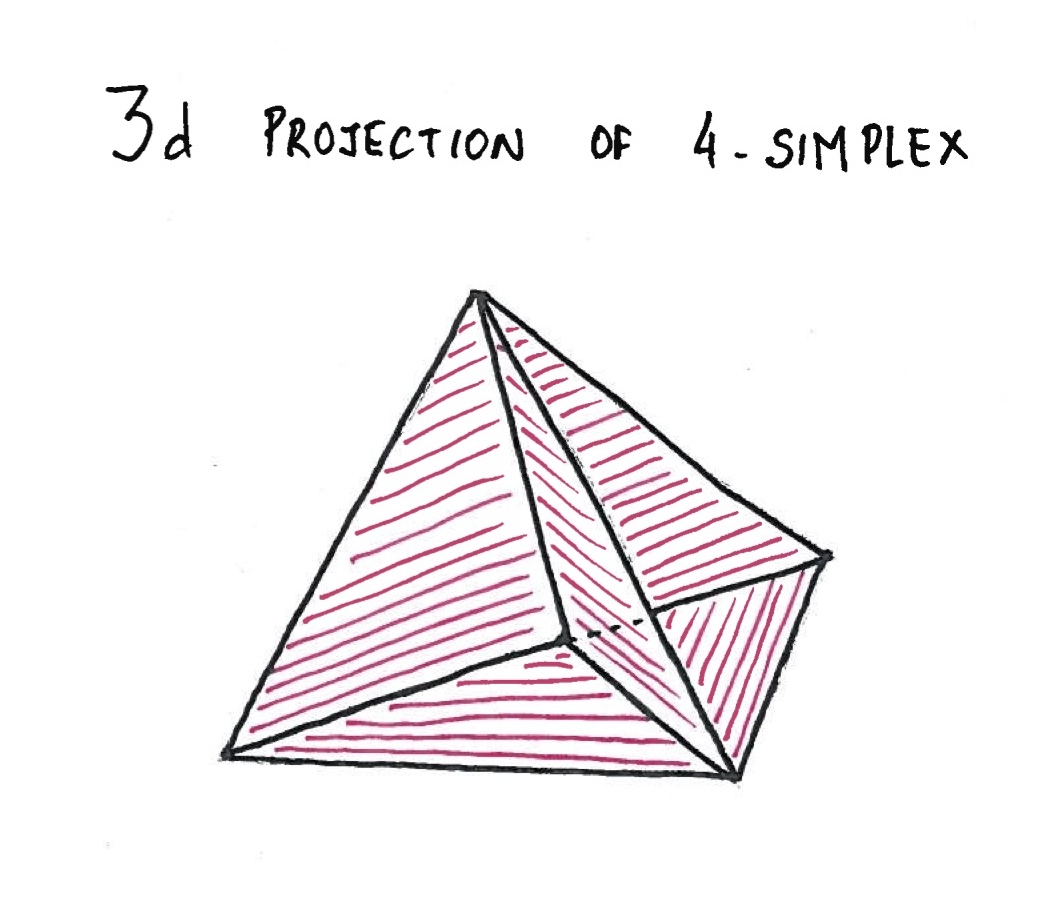}
\caption{3 dimensional projection of 4-simplex. One large tetrahedron composed of 5 tetrahedra,  4 inner tetrahedra plus the large one formed by all others.}
\label{4simplex} 
\end{figure}

The 3+1d construction is obtained by increasing the dimension of each of the structures in 2+1d. A time slice is now a 3d surface triangulated by tetrahedra (instead of the triangles in the 2d+1 case). Evolution moves, or Pachner moves are now represented by 4-simplexes. 4-simplexes are simply tetrahedra lifted up to four dimensions by adding a point in the extra dimension and drawing an edge to it from each node of the original tetrahedron. The difficulty of visualizing the 3+1d model lies in the fact that we can't draw tetrahedra in 4 dimensions but we can represent its projection in a 3d volume which, by analogy with the 2d projection of a 3d tetrahedron, is a tetrahedron composed of 4 internal tetrahedra composing 5 tetrahedra in total, see Figure~\ref{4simplex}.

4-simplexes in 3+1d have the same role as tetrahedra in 2+1d i.e. they represent Pachner or evolution moves. A 4 simplex takes one or more tetrahedra on a past space slice in 3d and evolve it to new tetrahedra in the future 3d slice. A four-simplex is composed of 5 tetrahedra; just as before we decide which Pachner move by selecting a few adjacent tetrahedra of the existing space foliation, placing the 4-simplex on top of those, erasing the existing tetrahedra and replacing them by the remaining tetrahedra in the 4-simplex. The centre of the tetrahedra is dual to a node and each triangle in the tetrahedra dual to links making up the four-valent spin network. 
 The possible Pachner moves\footnote{Generically a Pachner move in n+1 dimensions is generated by a n+1-simplex placed on top of $m=1,.., n$ existing n- simplices. The remaining n+1-m n-simplices become part of the new triangulation.} are then $m \rightarrow 5-m$, where $1\leq m\leq4 $.

\section{Energetic causal sets and causal spin foams}\label{ecs_intro}

An energetic causal set, is defined  \cite{ECS1,ECS2} is a causal set with additional intrinsic momenta labels.  These include $d+1$ dimensional momenta $p_a$ assigned to links, which are conserved on events. In fact there are two momenta associated to each link, $I<J$, an outgoing momenta from $I$, $p^{IJ}_a$ and an incoming momenta to $J$, labeled.
 $\tilde{p}^{IJ}_a$.

The conjugate quantities to the $p_a$ are space-time coordinates $x^a$ on links, these arise as lagrange multipliers which enforce constraints that ensure the conservation and flow of momenta through the causal set.  In particular, there are lagrange multipliers $z^a$ on events; these define an embedding of the events into a $d+1$ dimensional lorentzian space-time which emerges in the semiclassical limit.

In \cite{ECS1,ECS2} we proposed an action for an energetic casual set.
 \f
 S^0= \sum_I Z^a_I {\cal P}_a^I   +\sum_{(I,K)} (   X^{a I}_K {\cal R}_{aI}^K + {\cal N}^K_I {\cal C}^I_K +  \tilde{\cal N}^K_I  \tilde {\cal C}^I_K )
 + S^{int}
 \label{S0}
 \ff
 where the momenta are subject to three sets of constraints.
\begin{enumerate}

\item{}The first term in Eq.~[\ref{S0}] results from the conservation law associated with each event:
\f\label{conservation0}
{\cal P}_a^I = \sum_K p_{a K}^I -  \sum_L \tilde{p}_{a I}^L  =0 
\ff
where the sum over $K$ is over all events $I$ is connected to in the past and the sum over $L$ is over all events $I$  is connected to in the future.

\item{} The second term in Eq.~[\ref{S0}] comes from the redshifts constraint associated to each immediate causal link,  
\f\label{redshift0}
{\cal R}_{aI}^K = \tilde{p}_{aI}^K - U^{K b}_a  p_{bI}^K =0 
\ff
where $U^{K b}_a\in SO(d-1,1)$ is a parallel transport operator representing the time-like components of the space-time connection.

\item{} The third and fourth term result from the energy momentum relations for relativistic particles, two for each immediate link.
\f
{\cal C}^I_K = \frac{1}{2} \eta^{ab} p_{a K}^I p_{b K}^I  + m^2 =  0 , \ \ \ \ \  \tilde{\cal C}^I_K = \frac{1}{2} \eta^{ab} q_{a K}^I q_{b K}^I  + m^2=  0 
\ff

\end{enumerate}

The equations of motion include
\f
 Z^a_I -  U^a_b Z^b_K =  p^{a I}_K (\tilde{\cal N}_I^K  + {\cal N}_I^K )
 \label{emergence}
\ff
which can be interpreted as situating the events at points in an emergent space-time, separated by causal intervals proportional to the four momentum propagated between them, followed by rotation by a parallel transport operator.

An interesting feature is that if we take $S^{int}=0$, as we did in \cite{ECS1,ECS2}, the action is a pure linear combination of constraints.  
However we sill see below that we need a non-vanishing $S^{int}$ to represent a spin foam model.

The usual symplectic structure 
\f
\int x^a \dot{p}_a  ds
\ff
is gotten by considering a chain of events with each having a single input and a single output, as shown in \cite{ECS1,ECS2}.

We also gave a twistorial formulation of it for the null case when the masses  vanish.   
We represent null $p_a^{IJ}$ by a two component spinor $\pi^{IJ}_{A'}$ by
the correspondence
\f
p_a^{IJ} \leftrightarrow {\pi}_{A'}^{IJ} \bar{\pi}_A^{IJ} 
\ff
ie $p_a = \sigma_a^{A'A} {\pi}_{A'} \bar{\pi}_A $ for $3+1$ dimensional Pauli matrices, $ \sigma_a^{A'A}  $.

$q_a^{IJ}$  are similarly represented by spinors $\chi_A^{IJ}$.  
\f
q_a \leftrightarrow {\chi}_{A'}^{IJ} \bar{\chi}_A^{IJ} 
\ff
The redshift constraints Eq.~\ref{redshift0} are now
\f
{\cal R}_{AI}^K = \bar{\pi}_{AI}^K - U_A^B \bar{\chi}_{B I}^K =0 
\ff
The conservation law, Eq.~[\ref{conservation0}] at each event is
\f
{\cal P}_{AA'}^I = \sum_K \bar{\pi}_{A K}^I  { \pi}_{A' K}^I -  \sum_L \bar{\chi}_{A I}^L  {\chi}_{A' I}^L =0 
\ff

We again form an action only the energy-momentum relation constraints are not present because they are solved for
 \f
 S^{twistor}= \sum_I z^{AA'}_I {\cal P}_{AA'}^I   +\sum_{(I,K)}    \lambda^{A I}_K {\cal R}_{A I}^K  + \Omega_I^K  {\cal D}^I_K 
 \ff
 where there is a new constraint, fixing the helicity,
 \f
 { \cal D}=  \omega^A \bar{\pi}_A + \bar{\omega}^{A'} \pi_{A'}   -2S 
 \ff
 
 Let us consider the case that the redshift constraint is trivial, so that $U_A^B=I d_A^B$. In that case, we  can solve the redshift constraints and replace    $ \bar{\chi}_{B I}^K$ by  $\bar{\pi}_{AI}^K$.   The action reduces to
  \f
 S^{twistor}= \sum_I z^{AA'}_I {\cal P}_{AA'}^I   +\sum_{(I,K)}   + \Omega_I^K  {\cal D}^I_K 
 \ff
 
The variation of the action by $\bar{\pi}_{A I}^K$ yields the twistor incidence relation
\f
 \Omega^I_K  \omega^{A I}_K = z^{AA'}_I \bar{\pi}_{A' K}^I
\ff
\section{Wieland's twistorial spin foam action}\label{TSF}

We can now see that Wieland's twistorial action for spin foams\cite{WWnew,WW-hamiltonian} is very similar.

We construct a $3+1$ dimensional causal dual triangulation following Markopoulou's prescription described in section 2.  
Following Wieland's work we endow the elements of the dual triangulation, $T$, with the following degrees of freedom.

\begin{itemize}

\item{}We can begin with the original spin-network degrees of freedom, which are holonomies and fluxes assigned to edges of graphs embedded in spatial slices. Consider a link $\gamma$ of a spin network, $\Gamma$, which joins two nodes, which we may call 
$\gamma_i $ and $\gamma_f$.    To $\gamma$ we  can associate an initial flux 
$\Pi_\gamma \in sl(2,C)$, and a final flux, $\tilde{\Pi}_\gamma \in sl(2,C)$.  Conjugate to these are holonomy, 
$g_\tau \in SL(2,C)$.  

\item{} Dual to $\gamma$, in the triangulation of a spatial slice, is a triangle $\tau$.  In the spatial slice the triangle bounds two tetrahedra, these are each dual to one of the two nodes that $\gamma$ connects.  (Note that in the four dimensional simplicial complex there will be in general more tetrahedra bounding a given triangle)  

\item{} Thus to each triangle within a spatial slice we have
 an initial flux $\Pi_\tau \in sl(2,C)$, and  a final flux, $\tilde{\Pi}_\tau \in sl(2,C)$.  These represent the area of the triangle as seen from the frame of reference of each tetrahedra that bounds it.
 
\item{} Space-like parallel transport:
To each triangle in the tetrahedron, $\tau \in T $, dual in the 3-surface to space-like link, there is an holonomy, 
$g_\tau \in SL(2,C)$. They are related to the initial and final fluxes by the conservation constraint,
\f
{\cal R}^\tau_{ab} = \tilde{\Pi}^\tau_{ab} - ( g_\tau^{-1} \cdot \Pi_\tau \cdot g_\tau )_{ab} =0
\label{Rconstraint}
\ff
 
Note that as the triangles and the dual links to them are space-like, $g$ denotes space like components of the space-time connection. 

\item{} 
Similarly, each tetrahedron $T$ has associated to it two time-like four momentum, $p_a^T$ and  $\tilde{p}_a^T$ which are vectors in an internal momentum space, $\cal P$.  They correspond to total momenta incoming to, and total momenta outgoing from the tetrahedron from past and to future events.  They are parallel to the normals to the tetrahedra as seen in
the frame of reference associated with the two four-simplices that bound it, from the past and from the future.

\item{} Time-like parallel transport, or redshifts:

The $p_a^T$ and  $\tilde{p}_a^T$ are also  related by a parallel transport constraint
\f
{\cal R}_a^T = \tilde{p}_a - U(T)_a^b p_b =0 
\ff
where $U(T)$, which is an independent degree of freedom is, in this case, the parallel transport of the space-time connection in the time-like direction across the tetrahedron, from the event associated with the four-simplex that bounds it from the past to the event dual to the four simplex that bounds it to the future.

In the case that we can pick $U(T)=I$ we can solve ${\cal R}_a^T =0$ and equate $ \tilde{p}_a = {p}_a$

\item{}The four momenta and fluxes are related by three constraints:

The simplicity constraint, ensuring geometricity of the fluxes:
\f
{\cal S}^b_\tau = p_a^T \Pi^{ab}_{\tau} =0
\label{Sconstraint}
\ff
the volume shell constraint, establishing equivalence between mass, in the energetic causal set language, and volume,
\f
{\cal C}^T = p_a p_b \eta^{ab} + V^2_T (\Pi ) =0
\label{Vconstraint}
\ff
where internal indices are raised and lowered by $\eta^{ab}$, which is a metric on the internal momentum space
and $V_\tau (\Pi )$ is the volume of the tetrahedron.  We will see below why it is interesting to regard the volume to be
expressed as a four momentum. 

There is also the Gauss's law constraint, ensuring geometric closure of the tetrahedra
\f\label{gauss}
{\cal G}^i_T= \sum_{\tau \in T} \Pi^i_\tau =0
\ff
where $\Pi^i (\tau )$ is the dual, in the three-space orthogonal to $p^a$, of the flux $\Pi^{ab} (\tau )$.

The three volume of a tetrahedron, $T$ is a function of the fluxes across any three of its four triangles.
\f
V_T = \frac{\sqrt{2}}{3}\sqrt{|\epsilon_{ijk} \Pi^i (\tau_1 )\Pi^j (\tau_2 )  \Pi^k (\tau_3 )    |}
\ff

\item{}The spin network basis we used to represent the initial state is related to the $SL(2,C)$ connection representation in two steps.  First one transforms from the basis of $SU(2)$ spins, $j$ to the basis of $SL(2,C)$ representations, $(\rho, k)$ by the map
\f
j \rightarrow (\rho = \beta j, k=j)
\ff
Then one performs a non-linear fourier transform
\f
\Psi (\rho, k) \rightarrow \tilde{\Psi} (g) = \sum_k \int d\rho T_{(\rho, k)} (g) \Psi (\rho, k) 
\ff
Putting the two steps together we have,
\f
 \tilde{\Psi} (g) = \sum_k  T_{(\beta k, k)} (g) \Psi (\beta k, k) 
\ff

\item{} One solves the constraints (\ref{Rconstraint}) in terms of two twistors, 
\f
T_\alpha = (\omega^A , \bar{\pi}_{A'}), \ \ \ \  \ \ \tilde{T}_\alpha = (\tilde{\omega}^A , \tilde{\bar{\pi}}_{A'})
\ff
These are related to the holonomy and flux variables on the edge (or dual tetrahedra) by
\begin{eqnarray}
\Pi_{AB} &=& \omega_{(A} \pi_{B)}
\\
\tilde{\Pi}_{AB} &=& \tilde{\omega}_{(A} \tilde{\pi}_{B)}
\\
g_A^B & =& \frac{\omega_A \tilde{\pi}^B - {\pi}_A \tilde{\omega}^B}{\sqrt{\omega_A \pi^A}\sqrt{\tilde{\omega}_B \tilde{\pi}^B}}
\end{eqnarray}

These satisfy Poisson brackets
\f
\{\omega_A^\tau , \pi_{\tau'}^B \} = \delta_A^B \delta_{\tau \tau^\prime}
\ff
and similarly for the tilded twistor.  

The simplicity constraint (\ref{Sconstraint}) translates into a twisted helicity condition
\f
{\cal V}= \frac{1}{\beta +\imath} ( \omega_A \pi^A +  \tilde{\omega}_A \tilde{\pi}^A    ) + cc =0
\ff
and a linear constraint
\f
{\cal I} = p^{AA'} \omega_A \bar{\pi}_{A'} =0
\ff
\end{itemize}
The condition $\det g_A^B=1$ requires an area matching constraint on each triangle
\f
{\cal A}^\tau = \tilde{\pi}_A^\tau \tilde{\omega}^A_\tau -{\pi}_A^\tau{\omega}^A_\tau =0
\ff
Finally, the Gauss's law constraint, Eq.~\ref{gauss}, translates into
\f
{\cal G}^T_{AB}= \sum_{\tau \in T} \omega_{(A} \pi_{B)}=0 
\ff

We can now write the action for a causal spin foam as an example of an energetic causal set.  Note that the sum over events in the first term of Eq.~[\ref{S0}] translates into a sum over four-simplices, $I$, while the the sum over causal links in the second term translates into a sum over tetrahedra, $T$.  There is also also sum over triangles ensuring the associated redshift and simplicity constraints, Eqs.~[\ref{Rconstraint}] and [\ref{Sconstraint}],
 \f
 S^{csf}= \sum_I Z^a_I {\cal P}_a^I    +\sum_T  (   X^{a}_T {\cal R}_{a}^T + {\cal N}^T {\cal C}^T +  \tilde{\cal N}^T  \tilde {\cal C}^T  + A_T^{i}  {\cal G}^T_{i}) +\sum_{\tau} ( Y^{ab}_\tau {\cal R}_{ab}^\tau + u^\tau_a {\cal S}_\tau^a ) +
 \sum_{wedges} S^{wedge}
 \label{Scsf}
 \ff
where the constraint that energy and momentum are conserved at events, ${\cal P}_a^I=0$ becomes the condition that volume is relativistically preserved in Pachner moves, i.e. a sum over all tetrahedra, $T$ in the four simplex $I$,
\f
{\cal P}_a^I = \sum_{T \in \mbox{past set of I}} p_a^T - \sum_{T \in \mbox{future set of I}} p_a^T =0
\ff

\begin{figure}[t!]
\centering
\includegraphics[width= 0.5\textwidth]{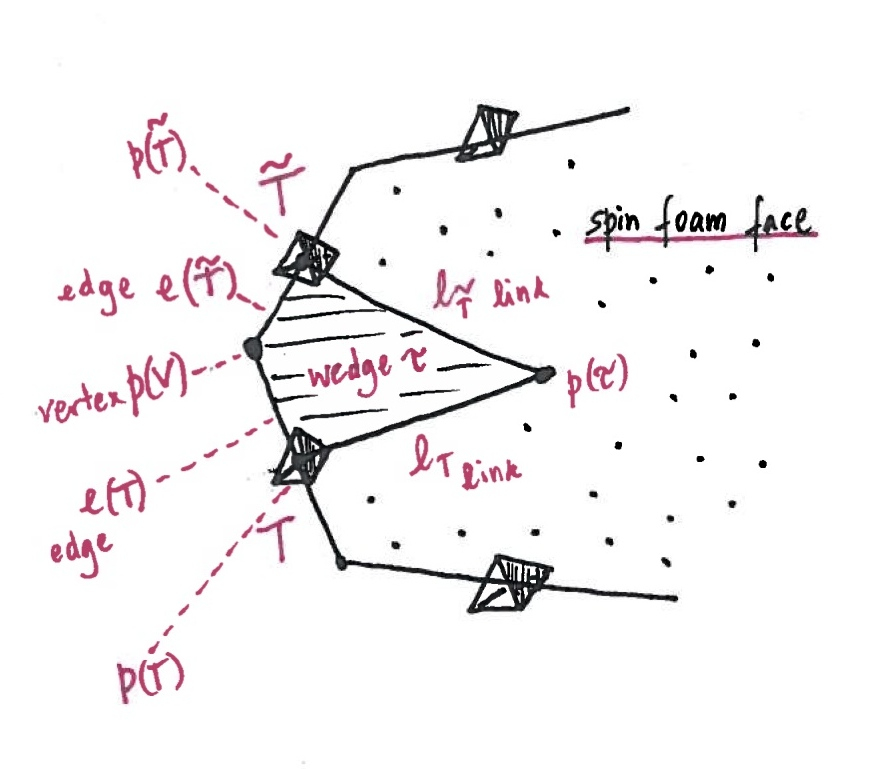}
\caption{Wedge integration in spin foam face. Figure courtesy of Wolfgang Wieland.}
\label{wedge} 
\end{figure}

The action Eq.~[\ref{Scsf}] is a sum of  constraints, plus the wedge term, $S^{wedge}$. Wedges divide the face in triangles for summation and are constructed as follows, see Figure~\ref{wedge}.  

A wedge is defined by selection of a triangle, $\tau$ and two tetrahedra it bounds within a given four simplex, $V^I$.  Let us call the two tetrahedra bounding $\tau$, by $T$ and $\tilde{T}$.   There will be a spin network link, $l_T$ joining a point, $p(\tau)$, on $\tau$ to the point $p(T)$ dual to, and within, $T$.  There is another link, $l_{\tilde{T}}$ joining the same  point, $p(\tau)$ on $\tau$ to the point $p(\tilde{T})$ dual to, and within, $\tilde{T}$.  Now dual to $T$ is a causal edge, $e(T)$,  joining $p(T)$ to the point  dual to and inside the four simplex, $V$.  Similarly the edge,  $e(\tilde{T})$,  joins  $p(\tilde{T})$ to $p(V)$.
The wedge is then the closed loop 
\f
w (\tau,T, \tilde{T},V)=  l_T^{-1}  \circ l_{\tilde{T}} \circ e(\tilde{T}) \circ e(T)^{-1}
\ff
The holonomy around $w$ is given by
\f
H (w)= h[ e(\tilde{T}) ] \circ h [e(T)^{-1}    ] \circ g (l)
\ff
The wedge action can then be taken from Eq.~[26] in Wieland's formulation of Hamiltonian Spin Foams \cite{WW-hamiltonian}:
\f
 S^{wedge}=-\frac{1}{2} M_w (h[ e(\tilde{T}) ] h [e(T)^{-1}    ])^{AB}
 ( \omega_A \tilde{\bar{\pi}}_B   + \bar{\pi}_A\tilde{\omega}_B         )+cc
 \ff
 where 
 \f
 M_\omega = \frac{1}{2} \left ( \frac{\sqrt{\pi \omega}}{\sqrt{\tilde{\pi}\tilde{ \omega}}}    +\frac{\sqrt{\tilde{\pi}\tilde{ \omega}}}{\sqrt{\pi \omega}}     \right )
 \ff

The spin foam partition function is then
\f
Z = \int \prod_{V_I} dZ_I \prod_T dX^a_T dN_T d\tilde{N}_T dp^T_a  d\tilde{p}^T_a d A_T^{i} \prod_\tau dY^{ab}_\tau 
du_a^\tau d\Pi_\tau d g^\tau_{A'}    e^{\imath S^{csf}}
\ff

The theory can be expressed in terms of twistor variables, in which case the action is
 \f
 S^{ctf}= \sum_I Z^a_I {\cal P}_a^I    +\sum_T  (   X^{a}_T {\cal R}_{a}^T + {\cal N}^T {\cal C}^T +  \tilde{\cal N}^T  \tilde {\cal C}^T +  A_T^{AB}  {\cal G}^T_{AB} ) +\sum_{\tau} (  \Omega {\cal V} + \rho {\cal I}+ \alpha_\tau {\cal A}^\tau)
 \label{Sctf}
 \ff
 and the partition function is 
\f
Z = \int \prod_{V_I} dZ_I \prod_T dX^a_T dN_T d\tilde{N}_T dp^T_a  d\tilde{p}^T_a dA_T^{AB}
\prod_\tau  d\Omega_\tau d \rho_\tau d\omega^A_\tau d\pi_{A'}^\tau d\alpha_\tau     
e^{\imath S^{ctf}}
\ff

 We can then apply to these spin foam models the mechanism for embedding a  causal set into  a Lorentzian space-time the method we developed for energetic casual sets\cite{ECS1,ECS2}.  Varying the action $S^{csf}$, Eq.~[\ref{Scsf}], by the momenta we find that the equations of motion fix intervals between the $Z_I^a$, which transmute from lagrange multipliers to coordinates of the events dual to the four-simplices. If $I$ and $J$ are two events causally linked through a tetrahedron, $T$, we have
 \f
 Z^a_J - U(T)^a_b Z^b_I = N^T p^a_T + \sum_{\tau \in T } u^\tau _b \Pi^{ab}_\tau
 \label{nonlinear}
 \ff
 
 Like Eq.~[\ref{emergence}], this can be interpreted as situating the events of the causal spin foam at points of an emergent four manifold coordinatized by the $Z^a_I$.  But this time the relation between the two causally related events is more complex,
 the momenta are time-like rather than null and the other factors may be indicative of curvature.   
 
 Again, the symplectic structure is gotten from a limit of a chain of events, as Wieland shows in \cite{WW-hamiltonian}  and as we show in \cite{ECS1,ECS2}.
 
Note that on any single edge we can choose to go to the analogue of $A_0=0$ gauge, which is $U=I$.

Then $\tilde{p}_a = p_a$, the redshift constraints can be dropped and the twistorial action simplifies to
\f
 S^{ctf}= \sum_I Z^a_I {\cal P}_a^I    +\sum_T  (   {\cal N}^T {\cal C}^T +  A_T^{AB}  {\cal G}^T_{AB}  ) +\sum_{\tau} (  \Omega {\cal V} + \rho {\cal I}+ \alpha_\tau {\cal A}^\tau)
 \label{Sctf2}
 \ff
while the partition function becomes,
\f
Z = \int \prod_{V_I} dZ_I \prod_T dN_T  dp^T_a  A_T^{AB} \prod_\tau 
 d\Omega_\tau d \rho_\tau d\omega^A_\tau d\pi_{A'}^\tau  d\alpha_\tau   e^{\imath S^{ctf}}
\ff
 The equation of motion from varying $p_a^T$ are
 \f
 Z^a_{T^+} -  Z^a_{T^-}= N_Tp^a_T + \sigma^a_{AA'}  \sum_{\tau \in T } \rho_\tau
 \omega_\tau^A \bar{\pi}^{A'}_\tau
 \ff
 where $T^{\pm}$ are the four simplices to the future and past of the tetrahedron $T$.

There are equations of motion from varying the $\pi^{A'}$'s and $\omega_A$'s.
\f
\bar{\omega}^{A'}_\tau (\alpha_\tau + \frac{\Omega_\tau}{\beta + \imath} ) + A^{A'B'}_\tau \bar{\omega}_{B'}^\tau +
\sum_{T | \tau \in T} ( \rho_\tau n^{AA'}_T \omega_A + 2 N V_T   \frac{\partial V}{\partial \pi^{A'}_\tau }     ) =0
\ff
Note that in terms of twistors
\f
V_T= \frac{\sqrt{2}}{3} \sqrt{|  (\omega^A \bar{\pi}_A )(\tau_1)(\omega^A \bar{\pi}_A )(\tau_2)(\omega^A \bar{\pi}_A )(\tau_3)  |  }
\ff

However, the $U=I$ gauge cannot be picked simultaneously on all the causal edges dual to tetrahedra, because there are in general multiple ways to connect an event to one in its causal future by future pointing sequences of causal edges.  By going forward on one such sequence and returning to the starting point by going backwards along another one forms closed loops, to which are associated gauge covariant holonomies.  These code information about the curvature of the space-time geometry.  
 
\section{Conclusion}\label{conclusion}

In a work appearing in parallel, Wieland \cite{WWnew} introduces a spin foam model which associates energy-momentum variables to the volume of tetrahedra in the spin network. These momenta are conserved in evolution moves, thereby introducing energy-momenta as fundamental variables of the model. Here we have shown that a closely related model can be associated to the energetic causal set we proposed in \cite{ECS1,ECS2} thus establishing a correspondence between this spin foam model and energetic causal sets. This endows the spin foam with a causal structure of its nodes and allows for the mechanism whereby space-time emerges in ECS to be considered also in the context of spin foams. 
This work suggests a new strategy for deriving a classical space-time from the semiclassical limit of the spin foam model, which makes use of the dynamically generated causal structure coded into every quantum history.

\section*{Acknowledgements}

We are grateful first of all to Wolfgang Wieland for many discussions and correspondence, and for sharing with us the results of his work.  M.C.\ was supported by EU FP7 grant PIIF-GA-2011-300606. This research was supported in part by Perimeter Institute for Theoretical Physics. Research at Perimeter Institute is supported by the Government of Canada through Industry Canada and by the Province of Ontario through the Ministry of Research and Innovation. This research was also partly supported by grants from NSERC, FQXi and the John Templeton Foundation.

\end{document}